\documentclass[preprint,12pt]{elsarticle}

\usepackage{amssymb}
\usepackage{amsmath,amssymb,amsfonts}
\usepackage{algorithmic}
\usepackage{graphicx}
\usepackage{textcomp}
\usepackage{xcolor}
\usepackage{paralist}
\usepackage[T1]{fontenc}
\usepackage{caption}
\usepackage{subcaption}
\usepackage{multirow}
\usepackage{booktabs}

\newcommand{\eg}{e.g.,\ }

\journal{Blockchain: Research and Applications}

\begin{document}

\begin{frontmatter}



\title{SmartQC: An Extensible DLT-Based Framework for Trusted Data Workflows in Smart Manufacturing}


\author[inst1]{Alan McGibney}

\affiliation[inst1]{organization={Nimbus Centre},
            addressline={Munster Technological University}, 
            city={Cork},
            country={Ireland}}

\author[inst1]{Tharindu Ranathunga}
\author[inst1]{Roman Pospisil}


\begin{abstract}
Recent developments in Distributed Ledger Technology (DLT), including Blockchain offer new opportunities in the manufacturing domain, by providing mechanisms to automate trust services (digital identity, trusted interactions, and auditable transactions) and when combined with other advanced digital technologies (e.g. machine learning) can provide a secure backbone for trusted data flows between independent entities. This paper presents an DLT-based architectural pattern and technology solution known as SmartQC that aims to provide an extensible and flexible approach to integrating DLT technology into existing workflows and processes. SmartQC offers an opportunity to make processes more time efficient, reliable, and robust by providing two key features i) data integrity through immutable ledgers and ii) automation of business workflows leveraging smart contracts. The paper will present the system architecture, extensible data model and the application of SmartQC in the context of example smart manufacturing applications.
\end{abstract}




\end{frontmatter}



\section{Introduction}

Data is increasingly becoming a strategic business resource that can eliminate existing bottlenecks in manufacturing lines and processes and disrupt traditional supply-chain models. Distributed Ledger Technology (DLT), including Blockchain, is seen as a key enabling technology to facilitate the open and trusted exchange of digital assets (data) over the Internet without using central servers or an independent, trusted authority. While industries are becoming more aware of the key benefits of DLT, such as security, immutability, availability, and transparency of data assets, the realisation and integration of these advanced digital solutions with existing business processes remains a significant challenge. Issues relating to interoperability, security, compliance, and data privacy are hindering the realisation of the next generation of smart factories.  

Advanced digital technologies such as the Internet of Things (under the guise of Industry 4.0), Machine Learning, Cloud Computing and Blockchain continue to play a significant role in the digital transformation of the manufacturing and industrial sectors. These enabling technologies are being leveraged to allow organisations to remain competitive, target the optimisation of existing business processes and navigate the digitalisation journey. The availability of consistent and reliable data using DLT creates significant scope for a manufacturer to introduce automated processes to collect, store, analyse and use production data that can, in turn, lead to an integrated approach to smart manufacturing. The rise of new digital economies (e.g. EU digital single market \cite{eu-single-market}) is revolutionising the way information is exchanged, shared, processed and analysed across industry sectors, new innovations and collaborative business models are emerging; however, this brings added complexities to already complex processes. For one, the sheer volume of data that is being generated demands a shift from centralised computing to decentralised processing. As such, traditional transactional models must be revised, and distributed data processing and storage architectures are being established to facilitate optimised data flows across disparate entities and stakeholders. Given the complexities in distributed data processing in novel digital ecosystems, there's a need to reevaluate transaction modelling to encompass and reflect the intricate properties of real-world manufacturing workflows. One of the most challenging aspects when developing such a solution is the selection and integration of specific DLT platforms that meet the requirements of specific application cases; as the technology matures and the number of platforms increases, it is becoming increasingly difficult for manufacturers to design and manage robust DLT networks to meet their specific needs. 

\subsection{Contribution}
The emergence of Blockchain technologies and solutions that are directly targeting the enterprise and industrial market has reinvigorated interest from industry in identifying the potential and real value this technology can bring to their specific businesses. As practitioners, this enthusiasm is welcome; however, there remains an air of caution among industry leaders as they try to navigate past the hype and make sense of the complexities of a new, yet rapidly evolving technology. Consequently, the main purpose of the work presented here is to provide insight into practical approaches that can support the industry in untangling the complexities of Blockchain for their needs and provide a framework that expedites the integration of enterprise-based DLT solutions into existing systems to create extensible and trusted data workflows. In particular, the proposed framework targets its application in quality control and product release through the trusted sharing of operational data. 

\begin{figure}[t!]
  \begin{center}
  \includegraphics[width=.8\textwidth]{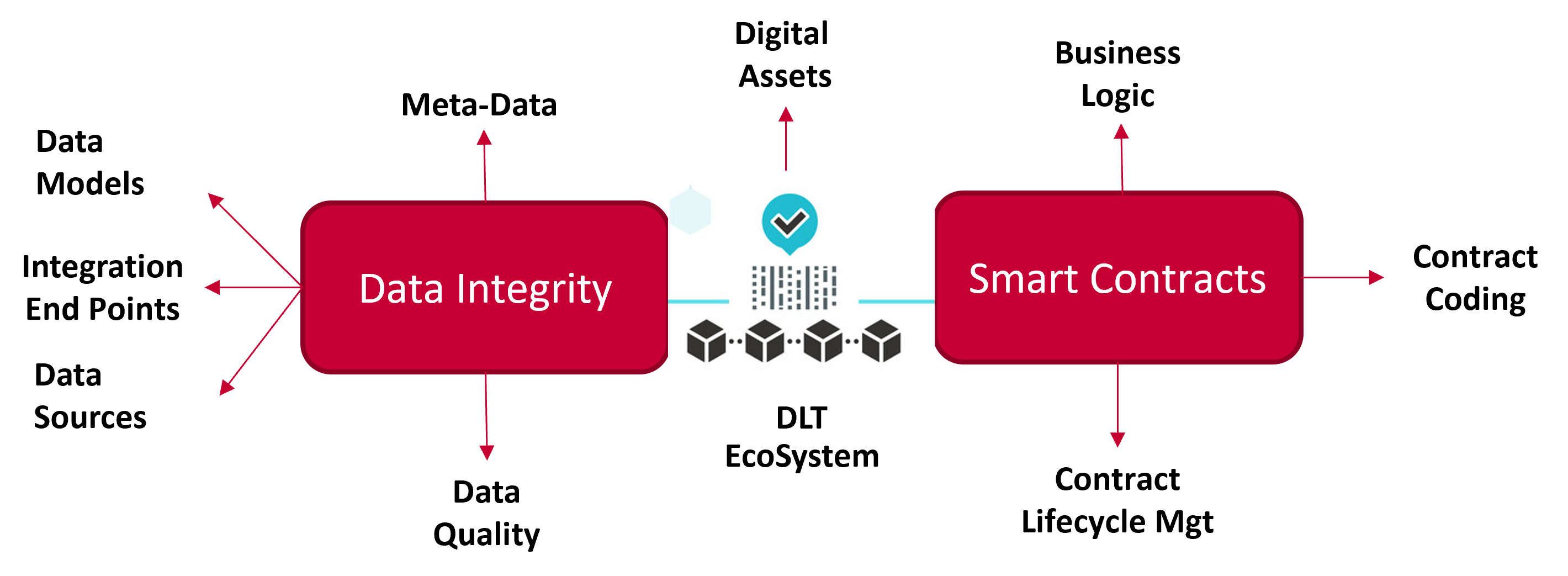}
  \end{center}
  \caption{SmartQC: A DLT-overlay for trusted data workflows}
   \label{fig:contributions}
\end{figure}

SmartQC is a DLT-based overlay for the deployment of trusted data-driven applications offering functionality, as presented in Fig. \ref{fig:contributions}. The following summarises the contributions of the work presented here:
\begin{enumerate}
    \item Analyse the role, opportunities and requirements for Blockchain technology in the context of smart manufacturing 
    \item Propose an architectural framework for integrating Blockchain as a trust layer for data-driven applications with an emphasis on data integrity and workflow automation. 
    \item Propose a novel transaction model that aims to effectively address ledger interoperability, complexities of distributed data processing and capturing relationships between digital assets. 
    \item The framework implementation is discussed as its application to smart manufacturing-specific use case scenarios and evaluation.
\end{enumerate} 

\subsection{Paper Organisation}
The remaining of the paper is structured as follows: Section \ref{sec:valueprop} outlines the value proposition and application case of the SmartQC platform, Section \ref{sec:framework} presents the SmartQC framework and architecture, Section \ref{sec:implementation} describes the current implementation, an initial evaluation of SmartQC components is provided in Section \ref{sec:evaluation} and Section \ref{sec:conclusion} provides a conclusion and insight into further research for SmartQC.
\section{Background and Value Proposition}
\label{sec:valueprop}
There are a number of application cases in the manufacturing domain that lend themselves to the use of DLT; the following provides a summary of these use cases and related work. \textit{Supply chain management} is viewed as an ideal application case for Blockchain solutions given the multitude of independent stakeholders and interactions involved, its ability to streamline integration, interoperability and provide an unprecedented measure of provenance among supply chain participants. It will also play a major role in regulatory compliance by providing traceability and independent auditability. By virtue of its consensus mechanism and distributed nature, Blockchain, as a DLT, can provide the following characteristics to enable future secure digital supply chains: traceability, transparency \cite{[3]}, stakeholder engagement and cooperation \cite{[4]}, supply chain convergence and digitalization, and common structures \cite{[6]}. The authors in \cite{[7]} adopted an integrated approach with the use of the Industrial Internet of Things (IIoT) enabled private Blockchain to track the location and status of raw materials in real-time. IIoT devices are deployed at each stage of the supply chain process to provide real time data and private Blockchain adding a layer of security and privacy to the architecture assuring not all the information is accessible by public. Supply chain actors working together will improve the overall efficiency of a Blockchain network \cite{[8]} However, it is has been acknowledged Blockchain should not be viewed as a complete solution that one can immediately adopt, the technology is still maturing, and its wider acceptance requires some core issues to be resolved such as data standardization, governance mechanisms, scalability, security and data privacy \cite{[11]}. 
\textit{Additive and Social Manufacturing (ASM):} DLT offers potential to support the protection of intellectual property (IP) in 3D printing for future manufacturing environments \cite{[15]}. It provides the ability to create a secure marketplace upon which 3D-printing designs and instructions could be exchanged, purchased or shared between parties while maintaining a degree of necessary control \cite{[13]}. It addition, it is possible to facilitate the sale and transfer of those designs using digital currency (self-executing smart contracts). Social manufacturing emphasises the ability to create highly customised and personalized products manufacturing driven by an ecosystem of individuals, companies, designers, artists, etc that participate in broader “maker communities”, using technologies such as 3D printing. The operation of the ASM currently involves the generation, gathering and storage of data in a centralized arrangement and therefore, data theft, manipulation and data biasing are considered to some of core challenges \cite{[14]}. 
\textit{Ensuring Validity and Secure logs of Critical Data:} The inherent properties of Blockchain platforms provides a mechanism to ensure tamper-proof storage and cross-verification of critical data, offering opportunities to detect and minimise the risk associated with fraud is explored in several scientific contributions \cite{[23]}. 
Trust in products through public data: Customers are becoming more cognisant of the impact the products they purchase has on the environment and sustainability of the processes used to manufacture them. Blockchain offers the ability to provide transparency of the process e.g. through ledger-based lifecycle assessment. By making this available through a public ledger (e.g. energy certification and use of renewables, sustainability labelling, traceability sustainably sourced raw materials) provides trust in the origin and lifecycle of the product. This aims to strengthen the customer relationship with the business and product production process \cite{[25]}.
\textit{Authentication and Access control of Smart Devices:} As IIoT brings connectivity to the shop floor it also brings added risk and increases the attack surface of manufacturing networks. The convergence of Information Technology (IT) and Operation Technology (OT) is a significant challenge to manage the complexity and risk of data driven processes. This is driving a zero-trust approach to devices on a manufacturing network. Blockchain can offer opportunity to create tamper-proof digital identities \cite{[27]} and the use of smart contracts to support authentication and access control of smart IoT devices that are rapidly proliferating and becoming integral to existing manufacturing processes \cite{[30]}. 
\textit{Improved Tracking of Maintenance Work:} To minimise downtime requires sufficient data to proactively manage maintenance issues and coordination between several different parties. Blockchain provides an interoperable, single-source ledger that all participants can consult to receive real-time updates. This can create a system that is better equipped to identify and monitor maintenance progress, identify inefficient contractors and cross-verify processes are followed. 
\textit{Quality Control towards Zero Defect Manufacturing and Product Approval and Release Process:} For each step of a production process, Blockchain can serve as an interoperable storage for necessary certifications, signatures, and quality checks. This allows manufacturers and suppliers to have more fine-grained oversight of the quality control process performed and their results. The ledger can then be used to aggregate the results and automate product approval process and will ultimately provide a tamper-proof record of how well a product (or component) passes certain quality checks. By implementing this oversight and creating uniformity, the Blockchain can contribute significantly to the evolution of manufacturing processes towards zero-defect manufacturing and minimise the effort required for auditing and certification of these products. 

Ravishankar et al, that outlined the limitations of blockchain for data integrity issues in terms of weak stability, high latency, and low-throughput as part of the European SUNFISH project \cite{[38]}. The authors have also proposed a database based on the characteristics of Blockchain that can solve the aforementioned issues for data integrity in a cloud computing environment. Hang and Kim \cite{[40]} proposed the use of blockchain technology by defining a smart contract-based application that can enhance the sensing data integrity of IoT device with resource constraints using a permissioned network (Hyperledger Fabric). Their analysis shows that the Blockchain is able to improve the data integrity process and has the potential for accommodating more resource-constrained IoT devices. 

Hang et al. \cite{[43]} proposed the use of Blockchain technology for the purpose of maintaining data integrity in the context of the agricultural domain. They developed a fish-farm platform that leverages Hyperledger Fabric and smart contracts to avoid the data manipulation and tampering. Various experiments were conducted to evaluate their proposed work in term of usability and efficiency. Omar et al. \cite{[44]} proposed the use of blockchain technology, specifically, the Ethereum based smart contracts to help maintain the data integrity for clinical trials. Their method uses Interplanetary File System (IFS) for the distributed data management which is used for highly-sensitive data and uses cryptographic hashes to make the data immutable. Hao et al. \cite{[45]} worked on the intrinsic properties of the Blockchain technology to improve the data integrity process. In general, they proposed the use of collaborative verification peers within the decentralized model that constitute inter- and inner-group consensus protocols. Their results show that the proposed decentralized model can enhance the data security, integrity, and verification process better than the existing works. Choi et al. \cite{[48]} proposed the use of a private Blockchain platform for data integrity in programmable logic controllers (PLC) in nuclear power plant environment. They employed the proof of monitoring concept to evaluate the data integrity in their method. Steinwandter and Herwig \cite{[49]} proposed the use of Blockchain for data integrity in the domain of pharmaceutical industry. Their system was built using Ethereum based smart contracts that can detect the manipulation, tampering, and backdating of data, accordingly. Their study also proposed future research guidelines for using data integrity and Blockchain in the context of patients’ data safety and intellectual property protection. Jamil et al. \cite{[50]} identified concerns over the data integrity issues in the context of pharmacology and drug supply chain. They proposed a Hyperledger Fabric based supply chain management that uses the specifically designed smart contracts to update the health and drug records in time-limited access. Kumar et al. \cite{[51]} conducted a similar study, i.e. data integrity issues in supply chain management, but in the context of manufacturing. They used Ethereum based smart contracts along with permission Blockchain integrated with ERC20 interface to secure the data related to manufacturing operations. Automation is a significant value add of the digital era, particularly in the industrial sector. Smart contracts are considered one mechanism to enable self-execution of specific logic based on trusted ledger data to support end to end automation throughout a product lifecycle.  The authors in \cite{[53]} explored the potential of Blockchain enabled smart contracts with the integration of IoT targeting the use case of smart hyperconnected logistics. Similarly, in \cite{[54]} the convergence of smart contracts and IoT device is demonstrated to ensure data immutability and public accessibility to temperature records while lowering operating costs in the pharmaceutical supply chain. Authors in \cite{[55]} discuss the possibility of smart contracts to simplify a variety of financial aspects, lowering administration costs dramatically in the supply chain process.

It is clear from the literature that DLT solutions are evolving on regular basis with new contributions in all verticals including smart manufacturing, it can be concluded that for each specific use case, there is a need to assess the various approaches, algorithm, and integration solutions available and that there is yet a general framework available to cover the range of complexities associated with the smart manufacturing domain. From a data integrity perspective Blockchain properties offer value for data driven services, however there is a need to create solutions that can be easily applied across a multitude of use case scenarios. The key benefits that can be provided by Blockchain technology relevant to the manufacturing sector use cases can be summarised in the context of the following functionality:
\begin{itemize}
\item \textbf{Data Integrity \& Coordination:} Provide a single source of truth (through the use of consensus algorithms and cryptography) that can be queried, scrutinised and analysed in an independent automated manner providing transparency to all stakeholders. Every entity will be able to access the same data, which was added to the ledger through autonomous validation (following governance rules and procedures). 
\item \textbf{Collaborative Trusted Ecosystem:} Decentralised consortium governance and transparency introduced by Blockchain-enabled systems create equal opportunities and aligned incentives for all entities can encourage collaborate approach to create more streamlined business processes. A manufacturing value-chain can consist of many different entities that rely on trusted relationships (and traditional business contracts), Blockchain can help reduce the risk through digital identities and more transparent record-keeping systems for suppliers, distributors, contractors, manufacturers, regulators etc.  
\item \textbf{Integration and Automation of Data Workflows:} Data driven processes require inputs from various systems and heterogeneous sources, a distributed ledger can act as a key enabler to support interoperability across a diverse set of stakeholders and information systems by providing a common data model for trusted data exchange. The coupling of DLT and smart contracts (e.g. for smart quality control applications), provides potential impact in terms of reducing production cycle time (e.g. automated checks and validation, smart alerting to highlight exceptions, prevent rejection of batches (continuous checking and feedback) and a step towards real time release testing through the control of process parameters, monitoring of attributes. 
\end{itemize}
 
The goal of the research presented is to define an extensible DLT based framework that can act as a trust overlay to existing manufacturing architectures. This will allow companies to take advantage of the benefits described above by providing tools and supports that simplify integration strategies for DLT with existing products and processes. 



\section{SmartQC Framework Overview}
\label{sec:framework}
Following the analysis of the various applications and proof of concept solutions in the manufacturing domain, as well as engagement with industrial stakeholders a number of key characteristics of the solution were derived that can be translated into a set of design requirements for the SmartQC solution, these include: 
a) ease deployment and integration with existing systems through reusable data models and application programming interfaces (APIs); b) be extensible and adaptable to facilitate a broad range set of use case scenarios, this will avoid the development of bespoke solutions through a common infrastructure; d) future proof with regards the shifting Blockchain landscape, i.e. to allow for inter-change of new DLT platforms with minimal disruption to business applications and c) support automation of processes and procedures through self-executing business logic (smart contracts). The following describes an overview of the SmartQC framework to address these requirements.


\subsection{Extensible Data Model}\label{DataModel}

\begin{figure}
\centering
\begin{subfigure}{.4\textwidth}
  \centering
  \includegraphics[width=.9\linewidth]{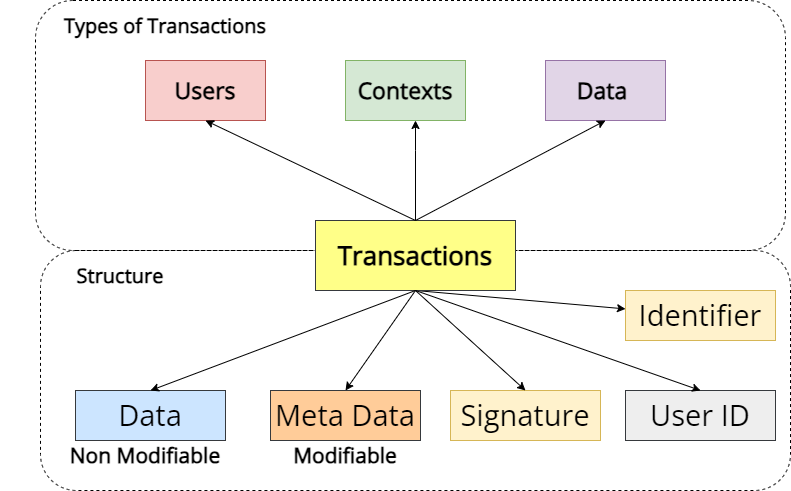}
  \caption{Generic Transaction}
  \label{fig:sub1}
\end{subfigure}%
\begin{subfigure}{.6\textwidth}
  \centering
  \includegraphics[width=.9 \linewidth]{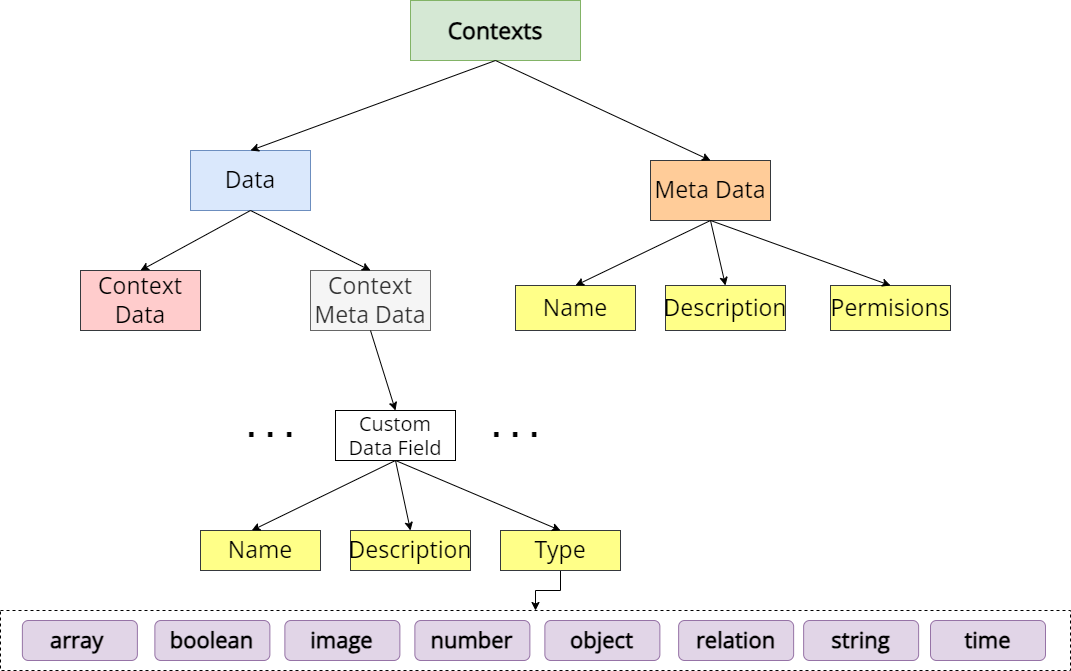}
  \caption{Context Transaction}
  \label{fig:sub2}
\end{subfigure}
\caption{Transaction Model}
\label{fig:transaction_model}
\end{figure}

To ensure the solution's extensibility, a scalable logical transaction model that maps entities, their relationships, and attributes, abstracting from the underlying characteristics of independent ledger layers is required. The result is a novel relational data model that facilitates a formal data schema for defining specific use case structures. This model empowers users to define custom scenarios involving transactions linked to multiple interrelated assets. Users can also specify constraints and data types within transactions, enabling fine-grained data validation before committing it to immutable storage (i.e., the ledger). Furthermore, the model provides an abstraction for enterprise applications, hiding ledger-specific data storage mechanisms. This abstraction eases implementation while offering scenario-agnostic CRUD-like operations on top of the ledger layer, similar to the asset registry layer proposed in \cite{[32]}. A populated data model is encoded as a \textit{transaction} in the SmartQC system, which defines three distinct transaction types:
\begin{itemize}
\item User Transaction - Encapsulates user-specific data (digital identifier and public key) to manage access control and authorization.
\item Context Transaction - Defines the format and contents of Data transactions and associated access rights. This transaction type is used to define the data structures of specific use case information.
\item Data Transaction - Represents instances of use case-specific data derived from the Context transaction.
\end{itemize}
A graphical representation of the SmartQC transaction model is depicted in Figure \ref{fig:transaction_model}. A generic transaction encapsulates four fields, summarized in Table \ref{tab:datamodel}.

\begin{table}[]
\tiny
\begin{center}
\begin{tabular}{@{}|l|l|@{}}
\toprule
Field & Description \\ \midrule
id & Unique identification of a transaction, mandatory field. \\ \midrule
data & \begin{tabular}[c]{@{}l@{}}Collection of key / value pairs such that each possible key\\ appears at most once, optional field.\end{tabular} \\ \midrule
metadata & \begin{tabular}[c]{@{}l@{}}Collection of key / value pairs such that each possible key\\ appears at most once, optional field.\end{tabular} \\ \midrule
signature & \begin{tabular}[c]{@{}l@{}}SHA3-256 hash of the transaction signed by Ed25519 \\ private key of an authorized user, mandatory field.\end{tabular} \\ \bottomrule
\end{tabular}
\caption{Generic structure of a transaction}
\label{tab:datamodel}
\end{center}
\end{table}

A transaction's unique identification is derived from the commit process in the relevant Blockchain. In each transaction, arbitrary data can be stored in the \textit{Data} and  \textit{Metadata} fields, with certain limitations intentionally enforced. The \textit{Data} field contains immutable data suitable for representing information that does not change over time. For example, in a manufacturing use case, the serial number for a specific device used in production can be stored in the \textit{Data} field. The  \textit{Metadata} field allows users to add additional data to a transaction, and its flexibility lies in being modifiable in each transaction over its life cycle. For instance, in a manufacturing scenario, \textit{Metadata} can be employed to track the production date and quantity of items manufactured in each transaction, allowing for dynamic updates as production continues. The generic transaction structure is extended to allow updated transactions, ensuring the traceability of all transaction modifications. An update transaction has the structure summarized in Table \ref{tab:update_transaction}.

\begin{table}[]
\tiny
\begin{center}
\begin{tabular}{@{}|l|l|@{}}
\toprule
Field & Description \\ \midrule
id & Unique identification of a transaction, mandatory field. \\ \midrule
asset\_id & Unique identification of the first create transaction, mandatory field. \\ \midrule
input\_id & Unique identification of the previous transaction, mandatory field. \\ \midrule
metadata & \begin{tabular}[c]{@{}l@{}}Collection of key / value pairs such that each possible key\\ appears at most once, optional field.\end{tabular} \\ \midrule
signature & \begin{tabular}[c]{@{}l@{}}SHA3-256 hash of the transaction signed by Ed25519 \\ private key of an authorized user, mandatory field.\end{tabular} \\ \bottomrule
\end{tabular}
\caption{Structure of a update transaction}
\label{tab:update_transaction}
\end{center}
\end{table}

The \textit{Data} field of \textit{User} transactions contains a username and the base58-encoded Ed25519 public key of the user. The \textit{Metadata} field is not generally required for a user transaction but can be useful in some use cases (e.g., editable data such as roles, organization details, etc.). Only administrators are authorized to perform operations on \textit{User} transactions.

\textit{Contexts} define the set of use case-specific transactions and how they are structured, akin to a database schema in a conventional relational database. As shown in Figure \ref{fig:transaction_model} (a), it inherits the generic transaction properties, and as shown in Figure \ref{fig:transaction_model} (b), it has its own hierarchy to represent its properties. The \textit{Data} field of \textit{Context} transactions defines the set of use case-specific transactions and how they are structured. It has the structure summarized in Table \ref{tab:context_data}. 

\begin{table}[]
\tiny
\begin{center}
\begin{tabular}{@{}|l|l|@{}}
\toprule
Field & Description \\ \midrule
context\textunderscore data & \begin{tabular}[c]{@{}l@{}}Collection of key / value pairs such that each possible key \\ appears at most once. Key must be equivalent to the referencing \\ data key. The values describe the content of the data with the same \\ key, optional field.\end{tabular} \\ \midrule
context\textunderscore metadata & \begin{tabular}[c]{@{}l@{}}Collection of key / value pairs such that each possible key appears \\ at most once. Key must be equivalent to the referencing metadata key. \\ The values describe the content of the metadata with the same key, \\ optional field.\end{tabular} \\ \midrule
version & Version object, optional field. \\ \bottomrule
\end{tabular}
\caption{Structure of a \textit{Data} field of \textit{Context} transaction}
\label{tab:context_data}
\end{center}
\end{table}

Context values have the base structure summarized in Table \ref{tab:context_values}. This basic structure of context values can be expanded according to the needs of the scenario, for example, by the format of the referencing data, order to sort data, or identification of searchable content value. The \textit{Metadata} field of the context transaction contains name, description, and permissions as a list of base58-encoded Ed25519 \cite{b9} public keys of users who have permission to insert or modify data transactions referencing this context. Only administrators are authorized to perform operations with context transactions.

\begin{table}[]
\tiny
\begin{center}
\begin{tabular}{@{}|l|l|@{}}
\toprule
Field & Definition \\ \midrule
name & Name of the referencing data or metadata, optional field. \\ \midrule
description & Description of the referencing data or metadata, optional field. \\ \midrule
\multirow{9}{*}{type} & Type of the referencing data or metadata, value must be one of \\ \cmidrule(l){2-2} 
 & "array", which describe array value, \\ \cmidrule(l){2-2} 
 & "boolean", which describe boolean value, \\ \cmidrule(l){2-2} 
 & "image", which describe image value, \\ \cmidrule(l){2-2} 
 & "number", which describe numeric value, \\ \cmidrule(l){2-2} 
 & "object", which describe object value, \\ \cmidrule(l){2-2} 
 & "relation", which describe related, parent, transaction, \\ \cmidrule(l){2-2} 
 & "string", which describe string value, \\ \cmidrule(l){2-2} 
 & “time”, which describe time value \\ \midrule
\multirow{8}{*}{content} & Array content definition, value must be one of \\ \cmidrule(l){2-2} 
 & "boolean", which describe boolean value, \\ \cmidrule(l){2-2} 
 & "image", which describe image value, \\ \cmidrule(l){2-2} 
 & "number", which describe numeric value, \\ \cmidrule(l){2-2} 
 & "object", which describe object value, \\ \cmidrule(l){2-2} 
 & "relation", which describe related, parent, transaction, \\ \cmidrule(l){2-2} 
 & "string", which describe string value, \\ \cmidrule(l){2-2} 
 & “time”, which describe time value \\ \midrule
parent & \begin{tabular}[c]{@{}l@{}}Identification of a related, parent, context transaction, \\ optional field. Valid only for "relation" type.\end{tabular} \\  \bottomrule
\end{tabular}
\caption{Context value structure}
\label{tab:context_values}
\end{center}
\end{table}

\begin{table}[]
\scriptsize
\begin{center}
\begin{tabular}{@{}|l|l|@{}}
\toprule
Field & Description \\ \midrule
id & Unique identification of a transaction, mandatory field. \\ \midrule
context\textunderscore id & Unique identification of a \textit{Context} transaction, mandatory field. \\ \midrule
user\textunderscore id & Unique identification of an authorized user, mandatory field. \\ \midrule
data & \begin{tabular}[c]{@{}l@{}}Collection of key / value pairs such that each possible key\\ appears at most once, optional field.\end{tabular} \\ \midrule
metadata & \begin{tabular}[c]{@{}l@{}}Collection of key / value pairs such that each possible key\\ appears at most once, optional field.\end{tabular} \\ \midrule
public\textunderscore key & Base58-encoded Ed25519 public key of an authorized user, mandatory field. \\ \midrule
signature & \begin{tabular}[c]{@{}l@{}}SHA3-256 hash of the transaction signed by Ed25519 \\ private key of an authorized user, mandatory field.\end{tabular} \\ \midrule
timestamp & Timestamp derived from the commit process in the relevant Blockchain, mandatory field. \\ \bottomrule
\end{tabular}
\caption{Structure of a create \textit{Data} transaction}
\label{tab:data_transaction}
\end{center}
\end{table} 

The \textit{Data} transaction has the base structure summarized in Table \ref{tab:data_transaction}. The structure of the data that goes into the ledger as a use case transaction can be defined within \textit{Contexts}, enforcing the constraints on data types and relationships. A summary of the transaction's data field definition is given in Table \ref{tab:context_values}, which is defined during the corresponding \textit{Context} creation. One of the major features here is the "relation" data type that can be included within a transaction. It enables linking multiple types of transactions together, representing multiple assets in the real world. It is useful in establishing intricate connections between various transactions, allowing for a more comprehensive and interconnected view of the assets and their interactions within the ledger. This capability is particularly valuable in scenarios where assets may have complex dependencies, such as supply chains, financial transactions, or asset management where complex queries for transaction retrieval are required. 

In the proposed transaction model, a crucial feature known as "conditional transaction creation" was implemented. This feature ensures that a new transaction can only be created if a specific prerequisite transaction already exists within the system. The enforcement of this condition occurs during the creation of a transaction context, where one context is associated with another by utilizing the "parent" field in the context structure. The concept is quite straightforward: before a new transaction can be added to the ledger, it must have a parent transaction in place. The parent transaction serves as the condition that needs to be met for the new transaction to be valid. If the parent transaction does not exist, the system will prevent the creation of the new transaction. This conditionality provides a powerful tool for ensuring the integrity and consistency of the ledger data. Such a feature can be extremely useful in various scenarios. For instance, in manufacturing, the order of operations in the production process is critical for product quality. By linking each production step to a preceding step as a prerequisite, you ensure that tasks are completed in the correct order.


\subsection{Architecture Specification}
\begin{figure*}[t!]
  \begin{center}
  \includegraphics[width=.9\textwidth]{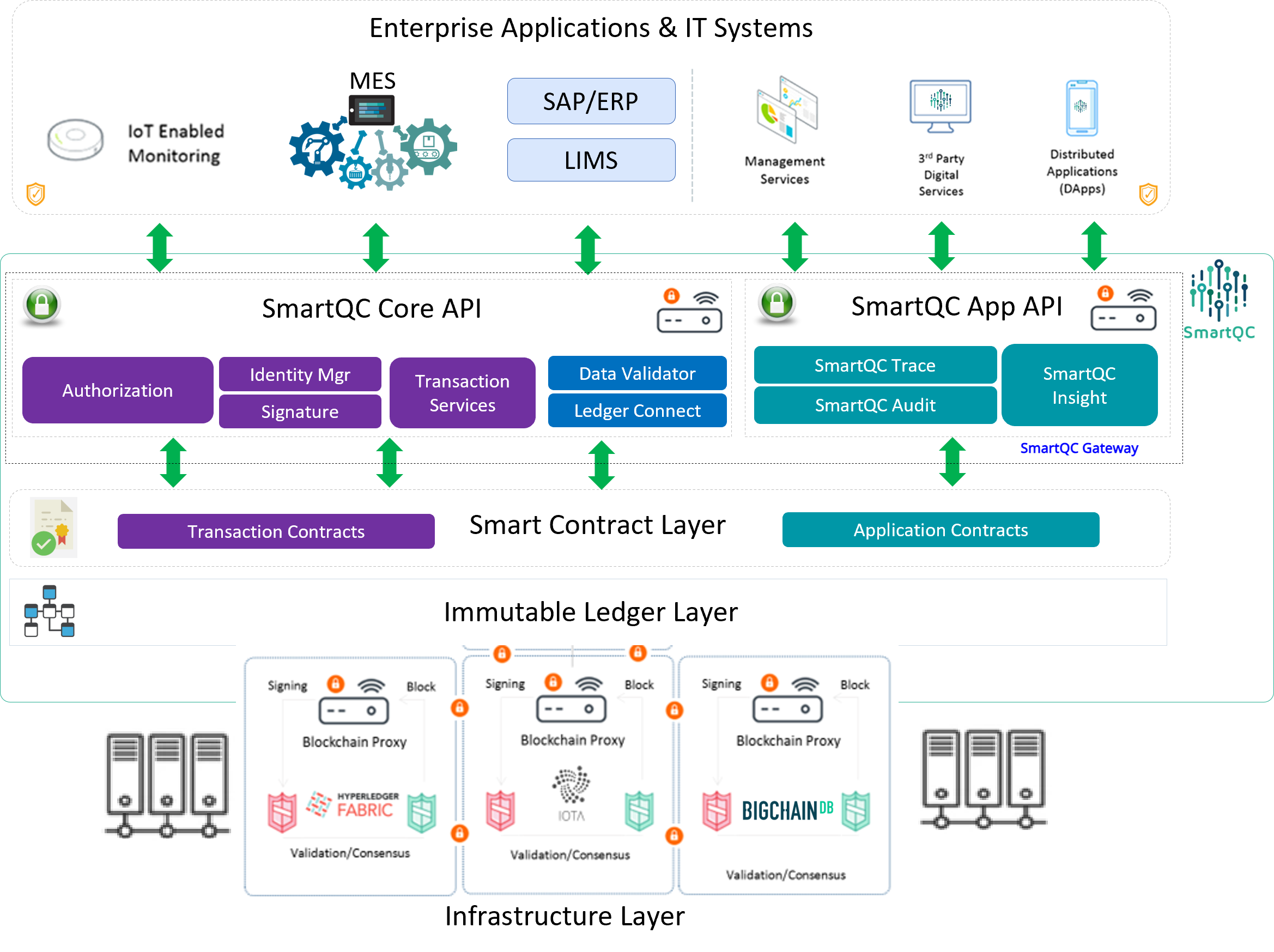}
  \end{center}
  \vspace{-1mm}
  \caption{SmartQC Architectural Layers}
  \label{fig:architecture}
\end{figure*}

The remaining requirements subsequently lead to the definition of a tiered architecture for DLT integration as depicted in Fig. \ref{fig:architecture}. It consists of 
four modular yet closely coupled layers: \begin{inparaenum}[(i)]
\item Enterprise Layer: This constitutes existing systems and data integration points, such as manufacturing execution systems (MES), Enterprise Resource Planning (ERP), Lab Information Management Systems (LIMS), field sensors and Internet of Things (IoT) devices. These end points are verified to be trusted (via registration and authorization) and utilise the secure SmartQC Restful application programming interface (API) to push or query data from the ledger layer. Moreover, this layer includes application specific support and administrative processes, such as user creation, identity management and data model configuration. Both web-based and mobile applications are provided to interact and integrate with SmartQC. 
\item SmartQC Gateway: Acts as a trusted oracle or intermediary gateway to the ledger layer. It provides an additional verification layer for digital signatures and data integrity. Moreover, it supports the interoperability between different distributed ledger/blockchain implementations through the integration of ledger specific connectors. Additional trusted gateways can also be deployed to facilitate the implementation of application specific business logic and functionality. This can be a customised set of API calls for a given scenario. 
\item Smart Contract Layer: Implements core contracts (e.g. authentication, transaction processing) and application contracts that are automatically triggered when certain conditions are met, these align with the application specific functionalities (e.g. automated release, certification, approval and confirmation of raw materials, products, etc).
\item Ledger Layer: provides the immutable storage for SmartQC transactions and transaction history using interchangeable distributed ledgers.  
\end{inparaenum} 
One of the key features of the proposed framework is the interchangeable ledger layer which supports interoperability. This allows the framework to leverage unique features offered by different DLT implementations depending on the application requirements. It is enabled by the novel data model to define transactions in a standard format, and the SmartQC gateway acts as the translator for different ledger implementations.  The following sections provide more details relating to the functionality of each of the architectural layers.

\subsection{Enterprise Layer}\label{Enterprise}
The enterprise layer represents the integration points of existing manufacturing components and services, from field devices up to business applications \eg{Quality Control, Order Management, Product release}. A SmartQC client library is provided for this integration. The \textit{SmartQC App API} provides a mechanism to enhance existing data workflows without replacing or impacting existing MES and ERP workflows, i.e. data is not duplicated rather only relevant data used for application specific needs is pushed to the ledger facilitated using the pre-defined use case context data definitions.

A set of user applications (both web and mobile) are provided to support administrative tasks supporting the transactions related to administrating data models: \begin{inparaenum}[(i)] 
\item User Creation and Retrieval; 
\item Context Creation, Update and Retrieval; 
\item Data Transaction Creation, Update and Retrieval.
\end{inparaenum} 




Moreover, the front-end administrative application acts as a visualization tool for the underline Blockchain networks. It visualizes the latest state of any assets \eg{Context, User or Data Transaction} and their transaction histories. It includes the links to the user who initiated (signed) the transaction to provide traceability. Transaction signing is performed by a custom JavaScript (JS) library implemented based on TweetNaCl high-security cryptographic library \cite{b8}, an open-source implementation of the Ed255519 \cite{b9} algorithm.

\subsection{SmartQC Gateway}\label{API}

The SmartQC gateway verifies all incoming data of the extensible model and transforms it to/from the format required by the DLT layer. This gateway was implemented as a RESTful Web Service using Jersey/JAX-RS library and deployed in Apache Tomcat Java HTTP web server environment. The corresponding HTTP API provides methods to PUT or GET all transaction types (\textit{User}, \textit{Context}, and\textit{Data}) to invoke data integrity checks, validation of permissions, context definition, and retrieval of state or historical data, as shown in the Table \ref{tab:httpapi}.

\begin{table}[t!]
\tiny
\begin{center}
\begin{tabular}{@{}|l|l|l|@{}}
\toprule
Method & Endpoint & Description \\
\midrule {PUT} & {/user} & Create or update a \textit{User} transaction. \\
\midrule {GET} & {/users/\{user\_id\}} & \begin{tabular}[c]{@{}l@{}}Get a list of \textit{User} transactions that match\\ the create transaction identification user\_id,\\ including changes.\end{tabular} \\
\midrule {GET} & {/users} & \begin{tabular}[c]{@{}l@{}}Get a list of all \textit{User} transactions, including\\ changes.\end{tabular} \\
\midrule {GET} & {/state/users/\{user\_id\}} & \begin{tabular}[c]{@{}l@{}}Get the state of a User transaction that\\ matches the create transaction identification\\ user\_id.\end{tabular} \\
\midrule {GET} & {/state/users} & Get a list of states for all User transactions. \\
\midrule {PUT} & {/context} & Create or update a \textit{Context} transaction. \\
\midrule {GET} & {/contexts/\{context\_id\}} & \begin{tabular}[c]{@{}l@{}}Get a list of \textit{Context} transactions that match\\ the create transaction identification context\_id,\\ including changes.\end{tabular} \\
\midrule {GET} & {/contexts} & \begin{tabular}[c]{@{}l@{}}Get a list of all \textit{Context} transactions, including\\ changes.\end{tabular} \\
\midrule {GET} & {/state/contexts/\{context\_id\}} & \begin{tabular}[c]{@{}l@{}}Get the state of a \textit{Context} transaction that\\ matches the create transaction identification\\ context\_id.\end{tabular} \\
\midrule {GET} & {/state/contexts} & Get a list of states for all \textit{Context} transactions. \\
\midrule {PUT} & {/transaction} & Create or update a \textit{Data} transaction. \\
\midrule {GET} & {/transactions/\{transaction\_id\}} & \begin{tabular}[c]{@{}l@{}}Get the \textit{Data} transaction that matches the given\\ identification.\end{tabular} \\
\midrule {GET} & {/transactions?asset\_id=\{asset\_id\}} & \begin{tabular}[c]{@{}l@{}}Get a list of \textit{Data} transactions that match\\ the create transaction identification asset\_id,\\ including changes.\end{tabular} \\
\midrule {GET} & {/transactions?context\_id=\{context\_id\}} & \begin{tabular}[c]{@{}l@{}}Get a list of \textit{Data} transactions that match\\ the given create context transaction identification\\ context\_id, including changes.\end{tabular} \\
\midrule {GET} & {/transactions?parent\_id=\{parent\_id\}} & \begin{tabular}[c]{@{}l@{}}Get a list of \textit{Data} transactions that are related\\ to the transaction with the given create transaction\\ identification parent\_id, including changes.\end{tabular} \\
\midrule {GET} & {/state/transactions?asset\_id=\{asset\_id\}} & \begin{tabular}[c]{@{}l@{}}Get the state of a \textit{Data} transaction that matches\\ the given create data transaction identification\\ asset\_id.\end{tabular} \\
\midrule {GET} & {/state/transactions?context\_id=\{context\_id\}} & \begin{tabular}[c]{@{}l@{}}Get a list of states for\textit{Data} transactions that match\\ the given create context transaction identification\\ context\_id.\end{tabular} \\
\midrule {GET} & {/state/transactions?parent\_id=\{parent\_id\}} & \begin{tabular}[c]{@{}l@{}}Get a list of states for \textit{Data} transactions that are\\ related to the transaction with the given create\\ transaction identification parent\_id.\end{tabular} \\
\midrule {GET} & {/state/transactions/search?text=\{text\}} & \begin{tabular}[c]{@{}l@{}}Get a list of states for \textit{Data} transactions that\\ contain the specified text.\end{tabular} \\
\bottomrule
\end{tabular}
\caption{HTTP API of SmartQC gateway}
\label{tab:httpapi}
\end{center}
\end{table} 

\begin{figure}[t!]
  \begin{center}
  \includegraphics[width=.8\textwidth]{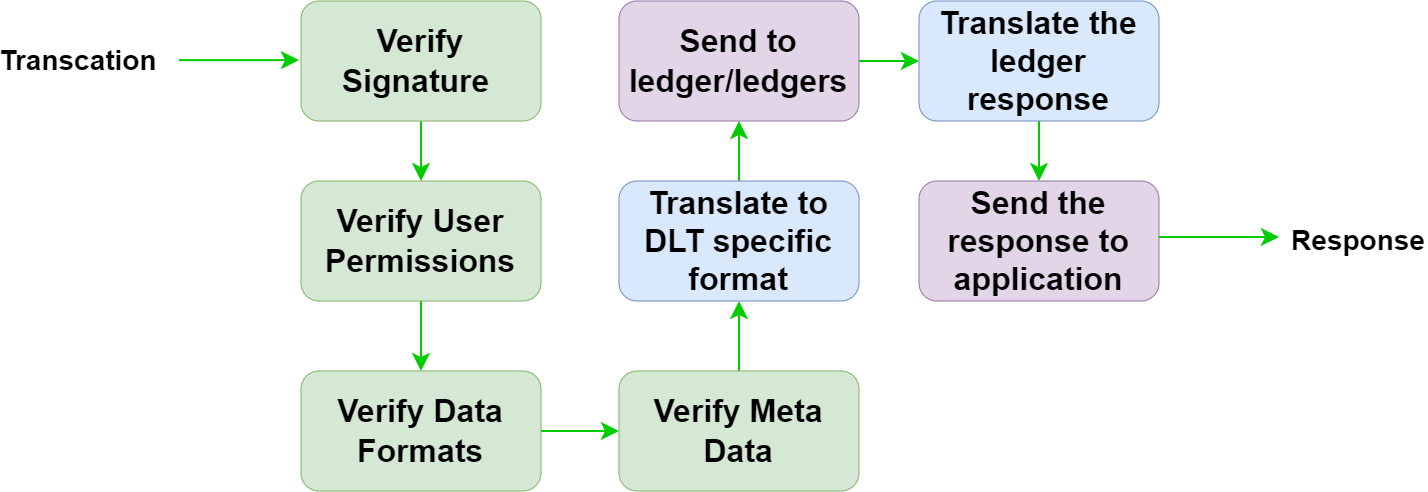}
  \end{center}
  \caption{SmartQC Gateway Workflow}
   \label{fig:gatewaylayer}
\end{figure}

These methods allow for the development of general applications for defining, configuring, managing and visualisation of any use case data model. Multiple SmartQC gateways can be distributed across participating organisations and services are deployed within trusted execution environments to ensure additional security and privacy. The workflow of the SmartQC gateway is shown in Fig. \ref{fig:gatewaylayer}. PUT user and context transactions are processed as follows:

\begin{enumerate}
\item Verify the signature of the transaction using the admin’s public key.
\item Verify the data and metadata structure.
\item Create a transaction in the format required by the DLT platform and forward it to the DLT layer end point.
\end{enumerate}

The process for PUT data transactions involves additional steps before committing to the ledger layer: 
\begin{enumerate}
\item Verify the signature of the transaction using the user’s public key.
\item Check user identity.
\item Check the related context transaction and permissions.
\item Verify that the data and metadata strictly conform to the definition given in the related context transaction.
\item Create arrays of parents and indexes to ensure searchability.
\item Create a transaction in the format required by the DLT platform and forward it to the DLT layer.
\end{enumerate}

This workflow ensures that all integrity checks permitted by the proposed SmartQC transaction model are validated and ultimately finalized as a failure or success. Overall, in the context of manufacturing, especially for Quality Control and Product Release scenarios, the SmartQC gateway emerges as a secure and effective method for managing and authenticating transactions, specifically tailored to these manufacturing use cases. The gateway's capacity to adeptly handle a variety of transaction types, each with customized processing procedures, underscores its adaptability and crucial role in upholding the integrity, security, and operational efficiency of the DLT system in data-rich applications.

\subsection{Ledger Layer}\label{Ledger}

The ledger layer in the SmartQC architecture serves as a pivotal component that connects various DLT infrastructures to the system. This layer is designed to accommodate multiple ledgers, enabling what is known as polyglot persistence \cite{fowler2011polyglot} in the application. The integration of multiple ledgers into the system is made possible by the SmartQC Gateway's ledger management component and the SmartQC transaction model. This feature empowers developers to create adapters that can seamlessly support the incorporation of diverse ledger technologies into the SmartQC ecosystem. In essence, the ledger layer provides the flexibility and extensibility needed to work with a range of blockchain solutions, making SmartQC a robust and interoperable platform for managing and processing distributed ledger data.

In the current implementation, the SmartQC gateway supports two types of DLT platforms: 
\begin{itemize}
\item BigchainDB~\cite{b10} a Blockchain database developed by BigchainDB GmbH. It was designed to combine blockchain properties (decentralization, Byzantine fault and Sybil tolerance, immutability, owner-controlled assets) and database properties (high transaction rate, low latency, indexing and querying of structured data). This combination of features makes it useful for a wide variety of use cases, including supply chain, IP rights management, digital twins \& IoT, identity, data governance and immutable audit trails. It uses Tendermint \cite{b11} for all networking and consensus. Each node has its own local MongoDB database \footnote{https://www.mongodb.com}, and all communication between nodes is done using Tendermint protocols.

\item Hyperledger Fabric \cite{b21} is an open-source platform initiated in 2015 for developing blockchain-based applications and solutions. Fabric is designed to be highly modular, scalable, and secure, making it ideal for use in enterprise-level applications. It provides a flexible architecture that allows developers to build custom blockchain applications tailored to their specific needs. It consists of a network of Peers, Orderers and Membership Providers organised in a permission-ed setting. The latest version of Fabric natively supports RAFT\cite{b22} consensus mechanisms within their ordering mechanism. This means it is able to provide Crash Fault Tolerance (CFT) to the SmartQC system.


\end{itemize}

\begin{figure}[t!]
  \begin{center}
  \includegraphics[width=.8\textwidth]{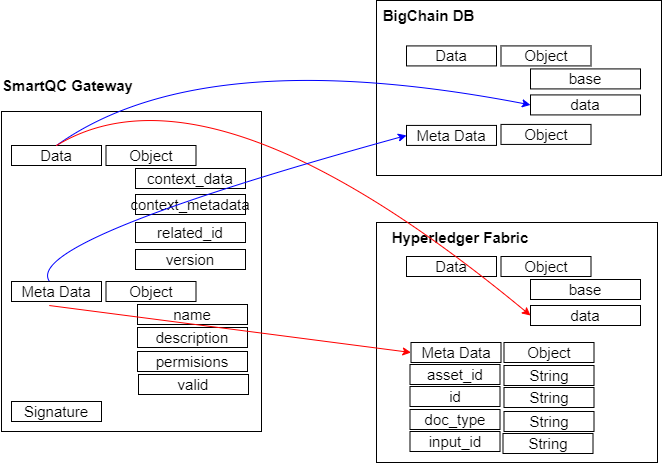}
  \end{center}
  \caption{Context data model mapping to BigchainDB and Hyperleder Fabric}
   \label{fig:transformation}
\end{figure}

These two platforms were chosen based on an evaluation of use case-level requirements. For instance, BigchainDB emerged as the preferred choice when relationships and owner-controlled assets are crucial due to its superior transactional modal; and for situations demanding the embedding of complex business logic, Hyperledger Fabric Chaincode was selected for its added flexibility. Apart from these two, the ledger platform IOTA \cite{b15} was considered as a candidate to implement into the SmartQC.  However, during the evaluation process, it was determined that the use of IOTA WASP was inappropriate, primarily because it lacks the capability to provide both a query against a state database and a history of key values over time. While querying the status database can be replaced by a less efficient solution at the plugin implementation level, the absence of a historical record represents a critical deficiency for the audit and transparency of the entire solution. 
One notable advantage of the SmartQC approach is the integration of these platforms through adaptors implemented within the SmartQC Gateway. These adaptors facilitate the transformation from a generic data model to the ledger-specific format with minimal intervention required. Consequently, the replacement of the ledger layer is anticipated to have a limited impact on the application-specific implementation and logic. 


The process of transaction transformation within the SmartQC ecosystem is a vital function facilitated through the utilization of a specialized SmartQC data model-specific smart contract. This pivotal component in the architecture plays a central role in ensuring seamless communication and data exchange between the SmartQC Gateway and the various ledger systems connected to it. The illustration provided, denoted as Fig ~\ref{fig:transformation}, exemplifies the intricate process of transaction transformation within the SmartQC ecosystem, a critical function that underpins the system's versatility and interoperability. This transformation is conducted through the utilization of a SmartQC-specific smart contract, designed to act as a translator between the SmartQC Gateway and the underlying ledger systems it interfaces with. This smart contract is more than just a mere mapping tool; it encapsulates the essence of the SmartQC's adaptability, enabling it to engage with diverse ledger platforms without necessitating substantial alterations at the application level. This transformative capability is underlined by the smart contract's ability to interpret and convert the generic data structures of the SmartQC model into ledger-specific formats. By doing so, it ensures that data integrity and the contextual relevance of transactions are preserved across different ledger environments. The smart contract processes elements such as context data, metadata, and signatures, ensuring that each transaction is not only compliant with the data model specifications but also with the requirements of each ledger system.

\section{Use Case Implementation}
\label{sec:implementation}
To provide a concrete example of how SmartQC is used in practice, an application for quality control to support manufacturing industry to move towards Zero Defect Manufacturing (ZDM), approval and product release was explored. For each step of a production process, SmartQC can serve as interoperable storage for necessary certifications, signatures and quality checks. This allows suppliers, manufacturers, auditors and customers have a clear oversight of the quality control process performed and their results. The ledger can subsequently be used to aggregate the results and automate the product approval process, and ultimately provide a tamper-proof record quality conformance of a product (e.g. feeding into a digital product passport). By implementing this oversight and creating uniformity, SmartQC can contribute significantly to the evolution of manufacturing processes towards zero-defect manufacturing and minimise the effort required for auditing and certification of these products. The deployment of the DLT network requires the definition of a minimal viable consortium (MVC). For the use case implementation the ecosystem aligns with a typical supply chain comprising of a number of independent stakeholders that interact and collaborate in a manner that potentially impacts the quality and management of a product. These organisations include: suppliers of raw materials, logistics partners, manufacturers (and business units such as procurement, technicians, automation, quality teams), third party machine calibration, distributors, consumers (customers), auditing and certification bodies. Each of these stakeholders may have specific roles within the DLT network defined by the governance procedure put in place, e.g. participants in the validation process, infrastructure provider or basic clients that interact with the ledger through authorised channels. The signing and verification for the final release is still very much a manual process. Digital tools have the potential to streamline the checking of the manufacture and verification of products in accordance with defined release procedures and the generation of certification of the finished product batch performed by a Qualified Person (QP). The objective is not replace all manual steps however by providing verifiable and trusted mechanisms for data integrity to signify that a product or batch is compliant with Good manufacturing practice (GMP) and the requirements of its marketing authorisation (MA) will lead to time savings for the QP and other actors (e.g. auditors) that require access to this information.


\begin{figure}[h]
  \begin{center}
  \includegraphics[width=.95\textwidth]{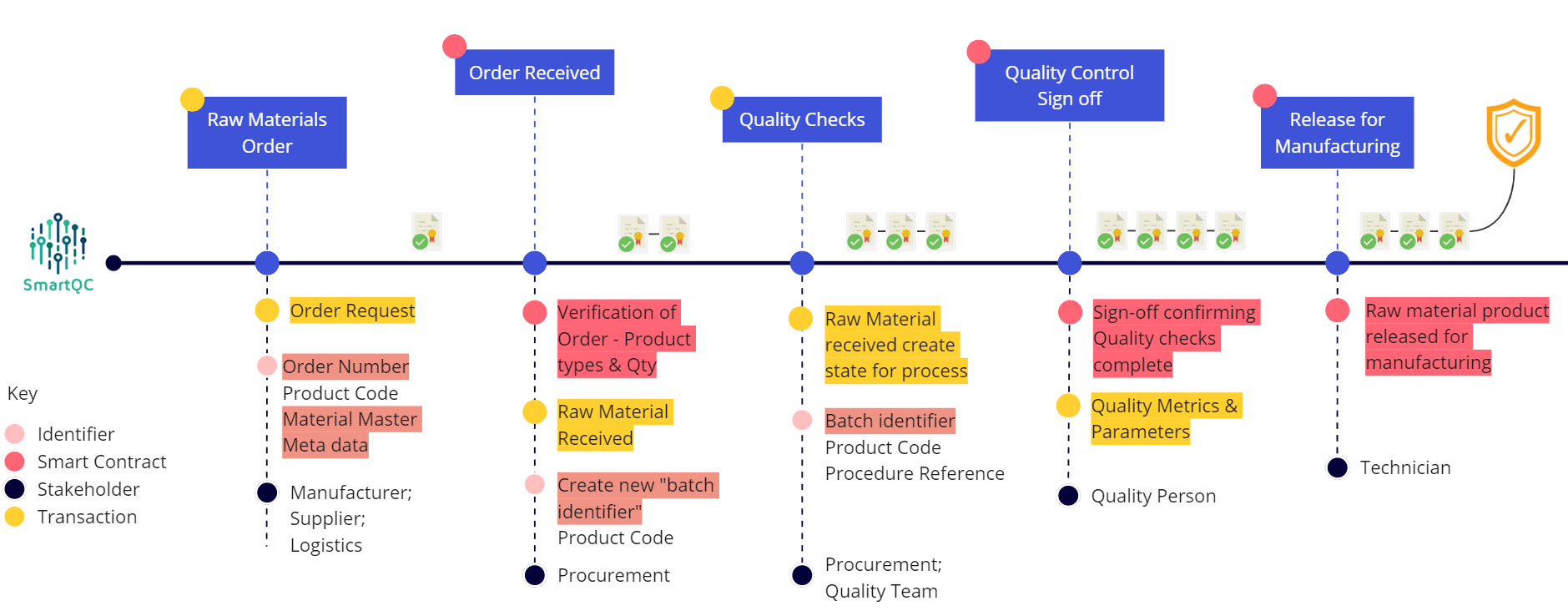}
  \end{center}
  \vspace{-1mm}
    \caption{Inbound Release Sequence and stakeholders}
   \label{fig:inbound_release_sequence}
\end{figure}

\begin{figure}[h]
  \begin{center}
  \includegraphics[width=.42\textwidth]{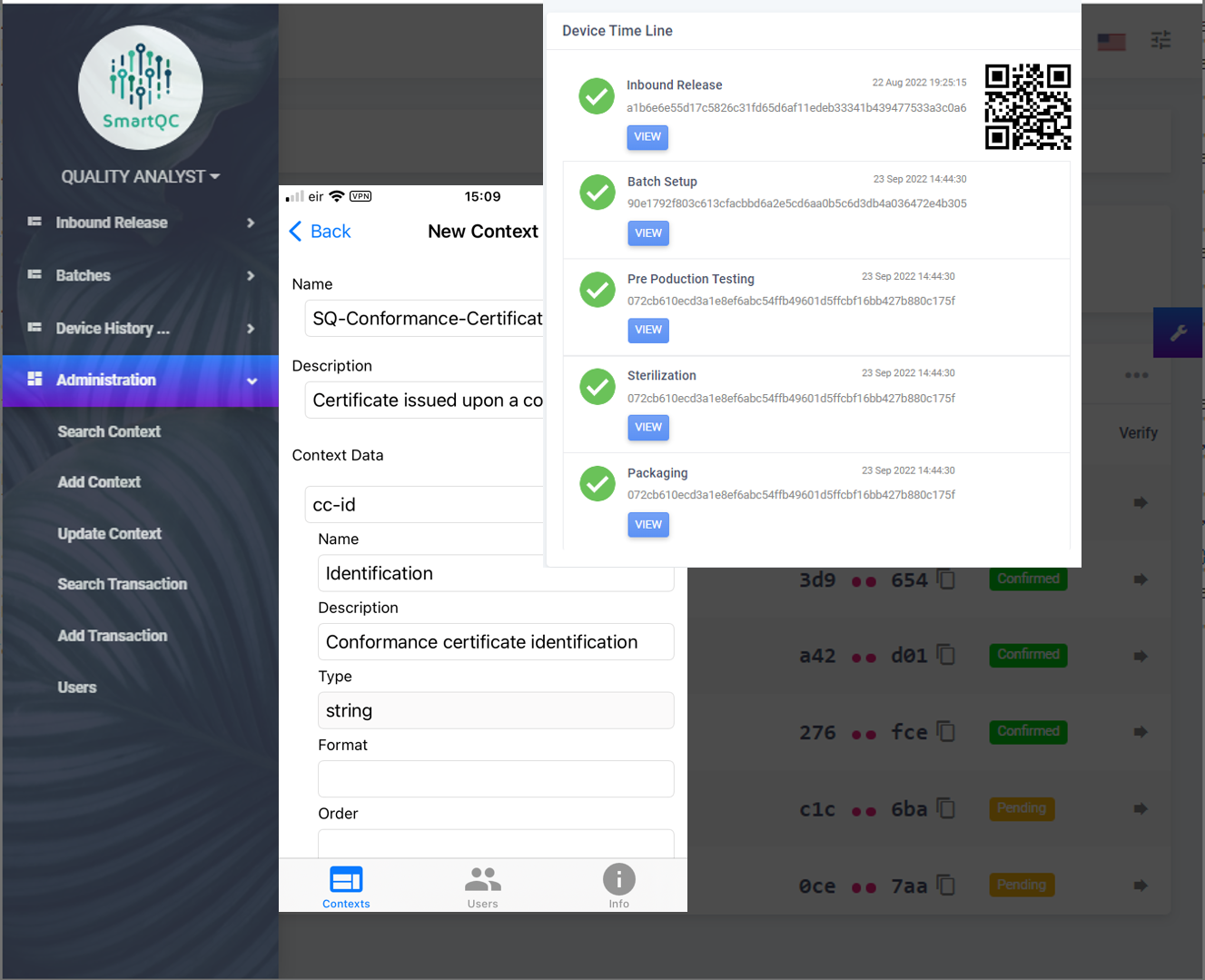}
  \end{center}
  \caption{SmartQC user applications and management tools}
   \label{fig:example_ui}
\end{figure}

For the proof of concept implementation the DLT networks (BigchainDB \& Fabric) are provided as a service to the involved stakeholders. A set of SmartQC nodes were configured and deployed on distributed cloud infrastructure to emulate geographically deployed instances aligned to relevant interacting organisations.

 Fig \ref{fig:inbound_release_sequence} describes an initial phase in the overall quality workflow, the receipt and processing of raw materials for the manufacturing process. This involves a procedure known as \textit{inbound release}, which includes manual and automated quality checks and verification of the raw materials prior to release to the manufacturing process. The initial step to configure this scenario on the SmartQC platform is the definition of the Contexts associated with the inbound workflow, these included for example context for i) orders ii) order lines, iii) material details, iv) quality procedures, v) quality checks and vi) conformance certificates. Fig. \ref{fig:example_ui} shows an example of user interfaces used to define contexts and demonstrate the inbound release functionality. Identifying data and metadata within these contexts is essential prior to deployment as only metadata can be changed with the data transactions, however SmartQC supports versioning of context data structures that allows for extension or new application cases. For instance, within the Quality Checks context definition the following data can be encapsulated: 

\begin{itemize}
\item Data  - Can include quality check IDs as referred by external systems and references to quality procedures that checks are based on. 
\item Meta Data - Array of quality checks consisting of properties that are being checked and their corresponding values, details of the sign-offs \eg{quality person, digital signature, time of the event and etc.}
\end{itemize}

 Fig \ref{fig:inbound_release_sequence} depicts an example sequence of actions taken during an inbound release as well as the stakeholder and type of interaction with the SmartQC ledger (data transaction, contract). Suppliers can receive order requests via the SmartQC platform, once ready for shipping they can commit digital quality conformance check that includes related data and sign-off. Once received at the manufacturing site additional checks and balances are carried out to verify the materials and cross check with its own procedures (e.g. valid suppliers encoded as smart contract, material cross check, visual inspection etc). All this data is posted as transactions to the SmartQC platform and stored on the immutable ledger layer. When the materials are brought to the manufacturing line they can again be verified with the full history or transactions, checks and conformance certificates available to the user. The steps of data checks, verification and sign-off can continue throughout the manufacturing process with final automated checks used to release the product for distribution. Data for these process steps are sent to the SmartQC gateway as data transactions (conforming to the context definitions. These transactions can be initiated from existing systems (through webhooks or alternative methods) and by actors (quality persons, technicians and etc.) within the system. The responsible person digitally signs these transactions using their private keys stored within secure wallets in client applications. Thus, it provides fine-grained traceability for the actions performed on data during the entire process. 
 The above example demonstrates how the proposed generic data model and the SmartQC Core API can be used to implement a custom manufacturing workflow depending on a use case. The proposed extensible DLT framework provides an abstraction for users having little or no knowledge of the DLT implementations to model their requirements and leverage the potential of DLT. 

\section{Evaluation}
\label{sec:evaluation}

\begin{figure}[h]
  \begin{center}
  \includegraphics[width=.82\textwidth]{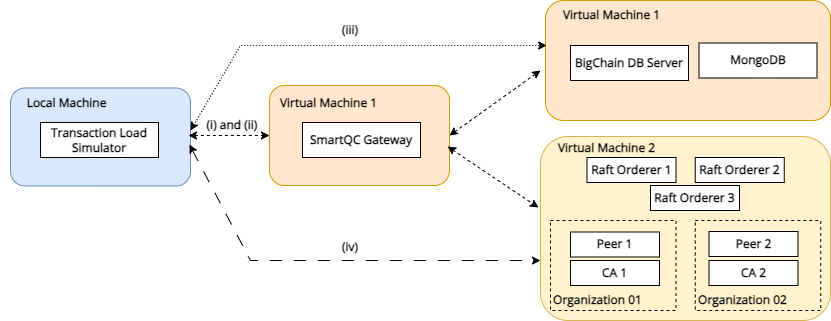}
  \end{center}
  \caption{System Under Test}
   \label{fig:sut}
\end{figure}

In this section, an evaluation of SmartQC is presented, focusing on its performance when integrated with BigchainDB and Hyperledger Fabric DLT platforms. This evaluation is crucial to understand how the SmartQC's impact on the latency and throughput of the transactions could directly impact existing manufacturing workflows. The evaluation is based on a benchmark conducted to evaluate the \textit{Latency}: the time taken from a transaction submitted by a client until a response is being received (round trip). The \textit{Throughput}: the number of total transactions that the system can handle was estimated based on the latency. The unit of measurement for the latency and throughput were milliseconds (ms) and transactions per second (tps), respectively. A custom service was developed using Javascript to generate the transaction load simulating a real-world system (eg: ERP or LIM) and to observe the round trip delay. 

\begin{table}[]
\scriptsize
\begin{center}
\begin{tabular}{@{}|l|l|l|@{}}
\toprule
\textbf{Parameter} & \textbf{Hyperledger Fabric} & \textbf{BigchainDB} \\ \midrule
Version & v2.1.0 & v2.2.2 \\ \midrule
Block Size & 10 transactions & N/A  \\ \midrule
Block Timeout & 250ms &  N/A\\ \midrule
Endorsement Policy & AND(’Org1.peer’, ’Org2.peer’) & N/A \\ \midrule
State DB & CouchDB & MongoDB \\ \midrule
Number of channels & 1 & 1 \\ \bottomrule
\end{tabular}
\caption{DLT configurations}
\label{tab:configs}
\end{center}
\end{table}

Four setup configurations were considered as part of the analysis and benchmarking, these include: (i) Endpoints of the SmartQC gateway connected to BigChainDB ledger, (ii) Endpoints of the SmartQC gateway connected to Hyperledger Fabric and (iii) BigChainDB endpoints bypassing the SmartQC gateway and (iv) Hyperledger Fabric endpoints bypassing the SmartQC gateway. During the (i) and (ii) configuration, all the endpoints that the SmartQC gateway exposes were benchmarked. During configurations (iii) and (iv) only the ledger read (GET) and write (PUT) were benchmarked, as ultimately every other endpoint of the SmartQC gateway is transformed into these two operations. Evaluations (iii) \& (iv) are used to demonstrate two integration patterns of SmartQC where the Gateway is connected to the ledger directly using the DB protocol (BigChainDB), and the ledger is connected through a smart contract (Hyperledger Fabric).

\textbf{System Under Test:} The benchmarks were conducted on a distributed deployment of the SmartQC, and it is depicted in Figure \ref{fig:sut}. The infrastructure used for the deployment is on the Google Cloud Computing Platform (GCP). As depicted in the figure, two virtual machines were used for the deployment considering the resource usage complexity of three software components, the SmartQC gateway and BigchainDB deployed on Virtual Machine 1(VM1) and Hyperledger Fabric on Virtual Machine 2 (VM2). The configuration of the two VMs is e2-standard-4 with 4 vCPU and 16 GB memory, and both the VMs are on a GCP local network and reside in the EU. The transaction load is generated on a PC that is connected to these instances over the internet. The specific configurations of the DLTs are summarised in Table \ref{tab:configs}. Using the custom load-simulating script, every endpoint was tested for 10 rounds with no concurrent transactions. In every round, latency was calculated, and the minimum, maximum and average latency were recorded for each endpoint. During the analysis, the average latency was considered to minimize the effect of outliers.

\begin{table}[]
\begin{center}
\tiny
\begin{tabular}{@{}|l|l|l|l|l|@{}}
\toprule
\textbf{SmartQC Operation} & \textbf{Type} & \multicolumn{1}{c|}{\textbf{Key}} & \multicolumn{1}{c|}{\textbf{\begin{tabular}[c]{@{}c@{}}Average Latency \\ with BigChainDB\\  (ms)\end{tabular}}} & \multicolumn{1}{c|}{\textbf{\begin{tabular}[c]{@{}c@{}}Average Latency\\ with Fabric\\ (ms)\end{tabular}}} \\ \midrule
/transaction & PUT & A & 71.98 & 3012.55 \\ \midrule
/users/\{user\_id\} & GET & B & 19.52 & 74.23 \\ \midrule
/state/users/\{user\_id\} & GET & C & 25.39 & 62.66 \\ \midrule
/users & GET & D & 54.32 & 322.78 \\ \midrule
/state/users & GET & E & 55.33 & 97.93 \\ \midrule
/contexts/\{context\_id\} & GET & F & 18.89 & 60.37 \\ \midrule
/state/contexts/\{context\_id\} & GET & G & 19.1 & 78.31 \\ \midrule
/contexts & GET & H & 187.63 & 1164 \\ \midrule
/state/contexts & GET & I & 194.02 & 112.16 \\ \midrule
/transactions/\{transaction\_id\} & GET & J & 16.64 & N/A \\ \midrule
/transactions?asset\_id=\{asset\_id\} & GET & K & 15.41 & 71.74 \\ \midrule
/state/transactions?asset\_id=\{asset\_id\} & GET & L & 19.34 & 85.28 \\ \midrule
/transactions?context\_id=\{context\_id\} & GET & M & 21.16 & 153.53 \\ \midrule
/state/transactions?context\_id=\{context\_id\} & GET & N & 23.92 & 77.91 \\ \midrule
/transactions?parent\_id=\{parent\_id\} & GET & O & 44.12 & 133.88 \\ \midrule
/state/transactions?parent\_id=\{parent\_id\} & GET & P & 54.47 & 88.51 \\ \midrule
/state/transactions/search?text=\{text\} & GET & Q & 22.32 & 76.28 \\ \bottomrule
\end{tabular}\caption{Latency of SmartQC Operations}
\label{tab:legend}
\end{center}
\end{table}

Table \ref{tab:legend} summarises the average latency of operations on the SmartQC when connected to BigchainDB and Hyplergeder Fabric. It was observed that the average latency for all the operations when SmartQC used BigchainDB stayed less than 200ms. In the case of Hyperledger Fabric, all the GET operations had an average less than 350ms. The transaction \textit{A} on Hyperledger Fabric had a much higher latency, this can be attributed to the involvement of the endorsement policy \footnote{https://hyperledger-fabric.readthedocs.io/en/latest/endorsement-policies.html} during ledger writes, which is specific to the Hyperledger Fabric protocol. However, these values are directly governed by the hardware configuration of the infrastructure and the specific DLT configurations, such as block size, block timeout etc. The individual values provide a baseline to understand the required configurations to support higher scalability scenarios. As of now, these latency levels are adequate for manufacturing quality control workflows, as there is no immediate requirement for near real-time validations for the given use case.

\begin{figure}
\centering
\begin{subfigure}{.49\textwidth}
  \centering
  \includegraphics[width=1\linewidth]{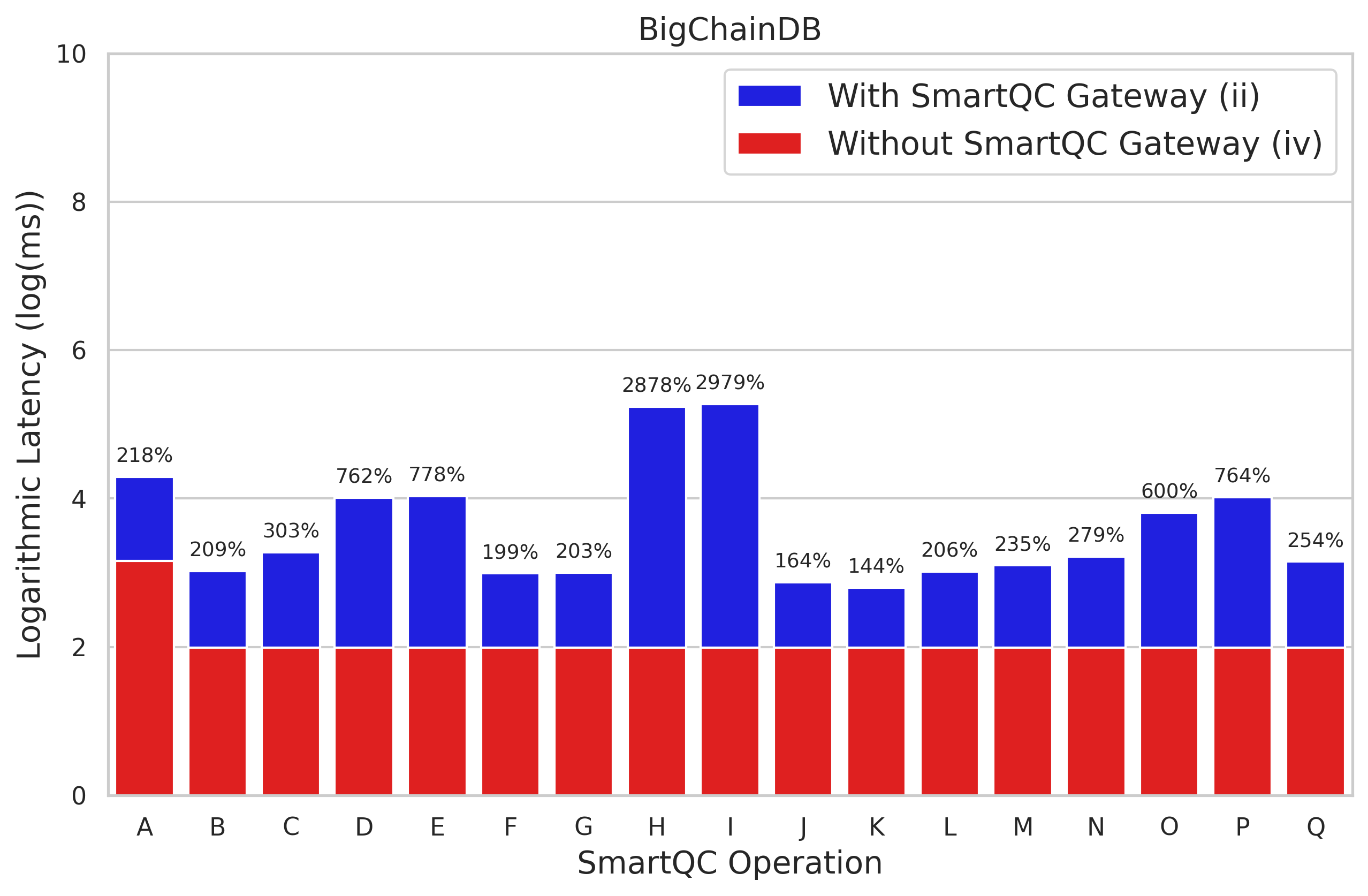}
  	\vspace{-7mm}
\end{subfigure}
\begin{subfigure}{.49\textwidth}
  \centering
  \includegraphics[width=1\linewidth]{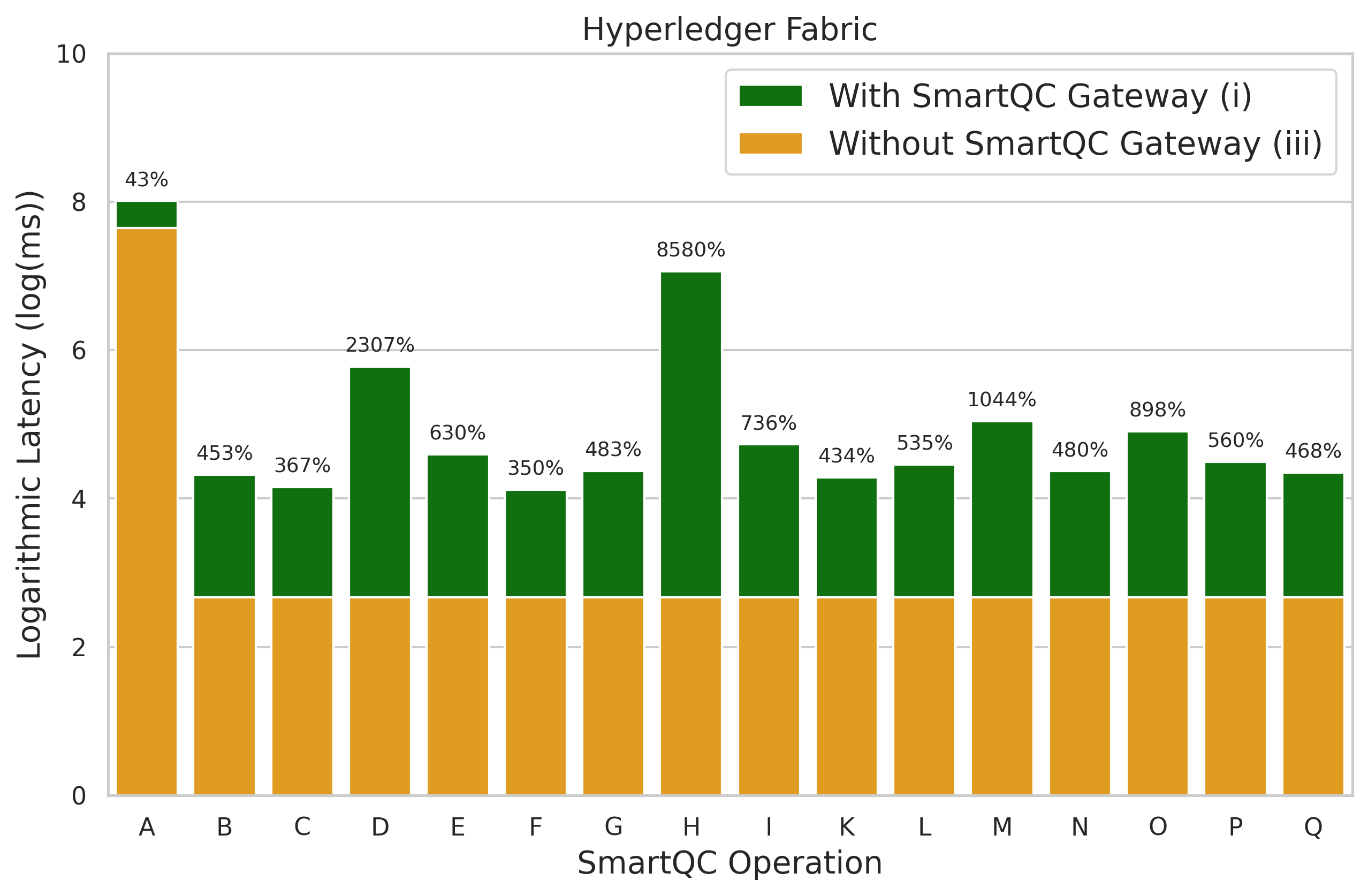}
  	\vspace{-7mm}

\end{subfigure}
\vspace{-1mm}
\caption{Overhead Latency Comparision}
\vspace{-5mm}
\label{fig:results}
\end{figure}

The primary goal of the benchmark was to analyze the SmartQC gateway's overhead, measurable in terms of latency, with the aim of enhancing the system's integrity and interoperability. In line with this objective, Figure \ref{fig:results} was created, deriving its basis from the benchmark results. It depicts the overhead latency caused by the SmartQC nodes for every operation that it offers. The objective is to extrapolate the conclusions of this for higher scalability scenarios. Both graphs in Figure \ref{fig:results} are on a logarithmic scale to bring out a clear visual representation of the differences in latencies (between ledger reads and writes), which might span several orders of magnitude. The logarithmic scale allows for a more straightforward comparison of the relative performance impact across various operations. 
To prevent any misunderstanding of the overhead represented on the logarithmic scale, each bar in the graph is labelled with the corresponding overhead percentage at the top.

As discussed in Section \label{API}  \textit{operation A} is significant interaction step in the manufacturing workflow as it commits data to the ledger. Moreover, it includes the extra validation processes introduced by the SmartQC gateway. In the context of BigchainDB  \textit{operation A} gets an overhead of around $218\%$ when using SmartQC gateway, and in Hypereldger Fabric, it is around $43\%$. This difference is caused by the nature of basic transactions offered by these two ledgers. In BigchainDB, transactions need to be arranged in a certain way before being added to the ledger, whereas Hyperledger Fabric has the flexibility of storing the transaction as it is, and this extra processing can incur processing delays. However, the main source of overhead is caused by the validation process of the SmartQC gateway, which validates the type of transactions, relationship with other associated transactions, data types, identities and signatures. This is really essential to ensure the integrity of data prior to committing to the immutable ledger (garbage in, garbage out). Moreover, it translates the transaction to the ledger-specific format, which enables multi-layer interoperability. Therefore, the overhead tolerance is a trade-off for achieving data integrity and interoperability between different ledgers.

The operations other than the \textit{operation A} are invoking ledger read transactions and translation of the transaction data in the proposed transaction model. There is some variability in how much latency is introduced by SmartQC Gateway, even for those SmartQC operations.  Specifically, \textit{operation H} and \textit{operation I} show a more substantial higher latency than others. This is because of the additional computation needed when creating the lists of context transactions and their states. In Hypereldger Fabric, the variability in latency is also present but tends to be less extreme than in BigchainDB. This is because Hyperledger Fabric supports the retrieval of transaction history using a native Chaincode operation, which is faster than BigchainDB.  

In conclusion, this evaluation sheds light on the latency overhead implications of integrating the SmartQC gateway with distributed ledger technologies (DLTs). The results, encapsulated in Table \ref{tab:legend} and Figure \ref{fig:results}, provide vital benchmarks for understanding the performance impacts in various ledger environments, extending beyond the specific cases of BigchainDB and Hyperledger Fabric. These insights are instrumental in gauging how SmartQC might perform with other types of ledgers, particularly in scenarios critical for data integrity and transaction validation.

The study highlights the inherent trade-off between the added latency due to necessary validation processes and the pursuit of data accuracy and interoperability across different ledger systems. This balance is crucial for optimizing system performance and maintaining reliability in diverse ledger contexts. The findings offer a valuable framework for future scalability and implementation efforts, ensuring effective integration of SmartQC in manufacturing quality control workflows, regardless of the underlying DLT.

\section{Conclusion}
\label{sec:conclusion}
DLT technology offers an opportunity to make compliance processes more time efficient and at the same time more reliable and robust through automation. The proposed SmartQC API driven architecture and extensible data modal has the potential to eliminate the need for paper trails using smart contracts that provide intrinsic evidence and provenance of information. The availability of consistent and reliable data using DLT creates significant scope for a manufacturer, including the decentralised management of production data to streamline Work order processing and tracking, Asset Lifecycle Tracking (conditions, PAT, alarms, events), processing of data-driven quality checks (Quality Metrics) and decision support tools. SmartQC aims to lower the barrier for manufacturers to embrace DLT technology providing for rapid prototyping across various use cases with SmartQC DLT overlay. Future work includes further benchmarking and preformance analysis of the solution and deeper investigation into the choice of individual ledger types to meet application needs. From an end user perspective the solution will be validated in the context of usability testing and ease of integration.

\section*{Acknowledgment}
This research has emanated from research supported by a research grant from Science Foundation Ireland (SFI) under Grant Number SFI/16/RC/3918 (CONFIRM)



 \bibliographystyle{elsarticle-num} 
 \bibliography{cas-refs}

\begin{thebibliography}{10}
\bibitem{eu-single-market}
t.~E.~E. Communication from the Commission to~the European~Parliament, the~Council, S.~Committee, the Committee of~the Regions, 2030 digital compass: the european way for the digital decade, 52021DC0118, COM/2021/118 final (2021).
\end{thebibliography}





\end{document}